\documentclass{ifacconf}

\usepackage{graphicx}      
\usepackage{natbib}        
\usepackage{xcolor}

\usepackage[T1]{fontenc}
\usepackage[utf8]{inputenc}

\usepackage{amssymb,amsmath,mathtools,epsfig}

\makeatletter
\let\old@ssect\@ssect 
\makeatother

\usepackage[ 
colorlinks=true,
pdftitle={Momentum-Based Learning of Nash Equilibria for LISA Pointing Acquisition},
pdfauthor={Aitor RG},
pdfsubject={LISA Pointint Acquisition},
]{hyperref}

\makeatletter
\def\@ssect#1#2#3#4#5#6{%
  \NR@gettitle{#6}
  \old@ssect{#1}{#2}{#3}{#4}{#5}{#6}
}
\makeatother

\usepackage{url}

\newtheorem{assumption}{\bf Assumption}[section]
\newtheorem{remark}{\bf Remark}[section]
\newtheorem{theorem}{\bf Theorem}

\usepackage[linesnumbered,ruled,vlined]{algorithm2e}
\SetKwInput{KwInput}{Input}                
\SetKwInput{KwOutput}{Output}              
\SetKwInOut{Parameter}{Parameters}

\renewcommand{\phi}{\varphi}
\newcommand{\reals}{{\rm I\!R}}

\begin{document}
\begin{frontmatter}

\title{Momentum-Based Learning of Nash Equilibria for LISA Pointing Acquisition}

\author[First]{Aitor R. Gomez}
\author[First]{Mohamad Al Ahdab}

\address[First]{Aalborg University,
    Aalborg, Denmark \\(e-mail: \{arg,maah\}@es.aau.dk).}

\begin{abstract}                
    This paper addresses the pointing acquisition phase of the Laser Interferometer Space Antenna (LISA) mission as a guidance problem. It is formulated in a cooperative game setup, which solution is a sequence of corrections that can be used as a tracking reference to align all the spacecraft' laser beams simultaneously within the tolerances required for gravitational wave detection. We propose a model-free learning algorithm based on residual-feedback and momentum, for accelerated convergence to stable solutions, i.e. Nash Equilibria. Each spacecraft has 4 degrees of freedom, and the only measured output considered are laser misalignments with the local interferometer sensors. Simulation results demonstrate that the proposed strategy manages to achieve absolute misalignment errors $<1\mu$rad in a timely manner.
\end{abstract}

\begin{keyword}
    Space exploration, Decision making and autonomy, High accuracy pointing, Extremum seeking and model free adaptive control,
    Game theory, LISA, Satellite constellation.
\end{keyword}

\end{frontmatter}

\section{Introduction}
\begin{figure}
    \centering
    \includegraphics{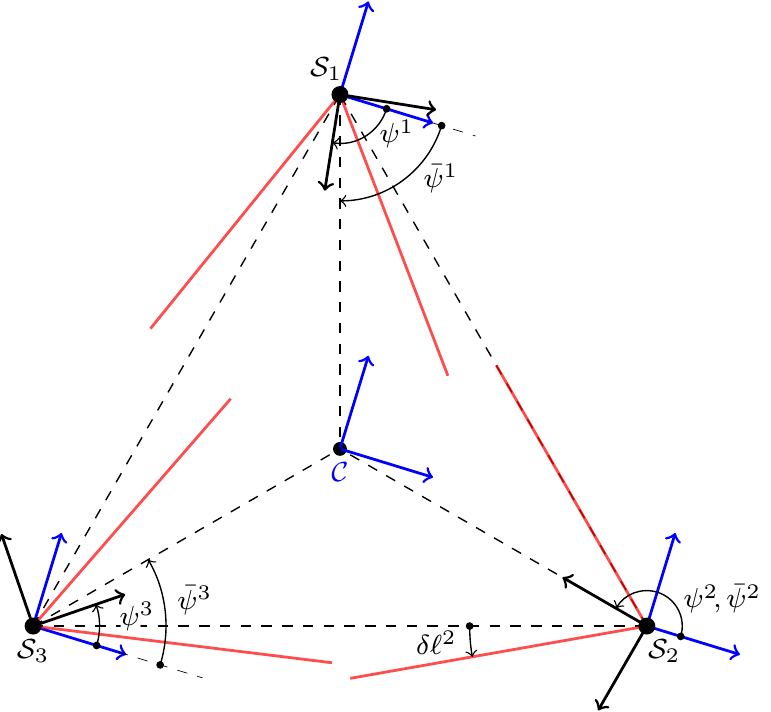}
    \caption{2-D geometry of the problem with overemphasized angles. It illustrates the constellation --inertial-- frame (blue), and the spacecraft' moving frames (black). The red lines represent the directions of the laser beams, $\psi^i$ the global state (yaw) and $\bar\psi^i$ the target global state (target yaw) of spacecraft $\mathcal{S}_i$.}
    \label{fig:geometry}
\end{figure}

The Laser Interferometer Space Antenna (LISA) is a highly complex space mission from NASA and ESA, aiming at detecting and characterizing gravitational waves (GWs). It is the result of the coordinated efforts of three spacecraft maintaining frequency-stabilized laser links between each other, resulting in an equilateral triangle formation as illustrated in Fig. \ref{fig:geometry}. Detection of GWs is enabled by the monitoring of frequency fluctuations of the lasers, produced by space-time ripples altering the distance between proof masses --the interferometer arm length-- located inside each spacecraft. The barycentre of the formation follows an heliocentric orbit lying in the ecliptic plane, with radius of 1AU and 20 degrees behind the Earth, and the interferometer arm length is $5\!\cdot\!10^6$km. This design has been meticulously chosen to allow for observations of gravitational radiation within a spectrum not reachable by its ground-based predecessor, the Laser Interferometer Gravitational-wave Observatory (LIGO). While LIGO is capable of detecting frequencies that are limited between $10$ Hz and $10$ kHz, LISA has the potential of complementing those observations by detecting GWs with frequencies ranging between $10^{-4}$ Hz and $1$ Hz.

\begin{figure}
    \centering
    \includegraphics{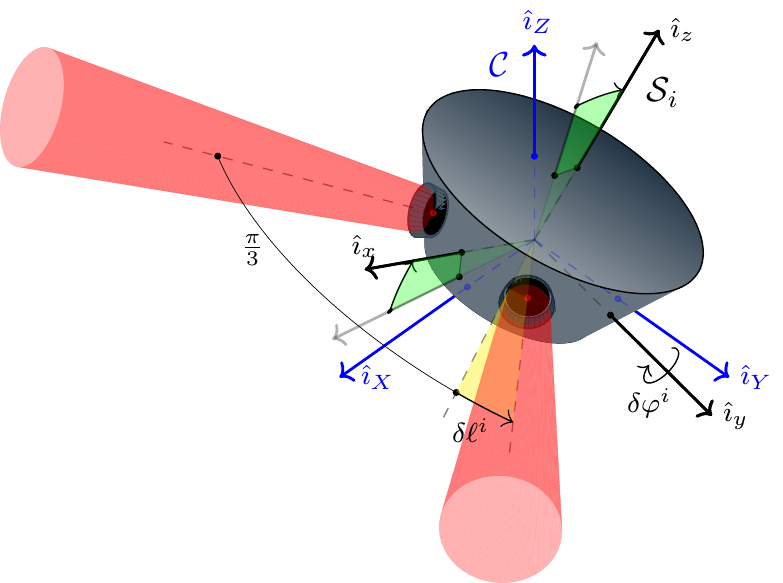}
    \caption{Constellation frame $\mathcal{C}$ (blue) and the spacecraft frame $\mathcal{S}_i$ (black). The green angle illustrates $\delta\phi^i$, i.e. an attitude change wrt its initial attitude (gray), and the yellow angle represents $\delta\ell^i$, a change in the telescope position wrt its default 60deg configuration.}
    \label{fig:spacecraft}
\end{figure}

The detection mode of GWs --or \emph{science mode}-- is the last of several intricate phases of the mission, most of which present very demanding requirements. The one studied in this paper is referred to as \emph{Pointing Acquisition and Tracking} (PAT) mode, namely, the task of aligning the local interferometer detectors with the lasers received from the other two spacecraft companions, while satisfying the remarkably low pointing tolerances required to engage science mode. To that end, every spacecraft is equipped with two identical optical assemblies (OAs) as payload, each of which contain, among others, the proof mass, a telescope, a laser source and laser interferometer detectors. The OAs have the telescopes pointing in a particular configuration that makes both of the lasers emitted by a spacecraft form an angle of 60 degrees by default. To account for the small drifting of the spacecraft throughout the heliocentric orbit that slightly deforms the equilateral formation, one of the telescopes is assumed to have an extra degree of freedom (see $\delta\ell^i$ in Fig. \ref{fig:spacecraft}). The laser interferometer detectors consist of a charge-coupled device (CCD) and a quadrant photodiode (QPD); the former used for signal acquisition with a field-of-view of 155$\mu$rad, and the latter for fine pointing with a field-of-view of $1\mu$rad. Precise measurements of the misalignments between the local and the received laser beams can be determined based on these sensor readings by employing the well-known method of differential wave-front sensing (\cite{Heinzel_2004}).

A preliminary pointing phase precedes PAT mode, in which Star Trackers (STRs) are first utilized to point the spacecraft. Due to multiple sources of uncertainty, the final pointing error of this phase is estimated in the \cite{LISA} technical report to be within a $9\mu$rad half-cone. Since the nominal laser beam is spanned by a smaller cone, two main strategies are usually considered \citep{OP_Rep} to deal with this uncertainty cone (UC) in the PAT phase: 1) defocusing of the laser beam to cover the UC  and 2) scanning of the UC. \cite{Maghami} developed an attitude control policy based on beam-defocusing, achieving the desired pointing accuracy, however, with no specification of time duration. \cite{Cirillo}, on the other hand, studied the scanning method and developed a control strategy to establish the laser links within a 70 min interval, approximately, for a 15$\mu$rad UC. Albeit, both strategies perform one link acquisition at a time.

In this paper, we propose an algorithm based on beam-defocusing for a simultaneous laser acquisition and fine-pointing, showing that fast and accurate convergence can potentially be achieved. As stated in the internal report \cite{LISA2}, defocus of the laser beam is, additionally, considered to be more robust than scanning. Note that simultaneity of these operations imply that all spacecraft must use their lasers and detectors at the same time. While the detection of the received laser is deemed nonviable if the local lasers are always on, this inconvenient could be avoided by emitting pulses, rather than a continuous laser wave. The repetition rate must be chosen to ensure receiving a pulse when the local laser is not emitting, and part of the travel time that each pulse takes to reach the opposite spacecraft --16 seconds for a 5$\cdot10^6$km interferometer arm length-- can be used to build up energy between pulses. At every detection of a pulse, the strategy described in this paper can be adopted to guide each spacecraft in unison. Alternatively, this methodology has the advantage of being combinable, by performing an initial scanning step to further reduce the UC and posteriorly defocusing the laser, or just as a high-precision pointing algorithm.

\subsection{Contribution}
The main contribution is Algorithm \ref{alg:NEseekings}, designed to provide a tracking reference to the spacecraft' controllers for a fast, simultaneous and accurate alignment of the laser beams into the interferometer detectors. It is based on the work by \cite{Tatjana_alg_2019} with the addition of two extra terms: the so-called momentum \citep{qian1999momentum} and residual-feedback \citep{Res_feedback_2022}. We propose to include these in the context of game theory in order to improve the learning rate of equilibria.

We show the performance of Algorithm \ref{alg:NEseekings} for the laser acquisition and pointing problem in the LISA mission introduced in the previous section. The problem has been formulated as a cooperative game enabled by the suggested pulsed laser strategy, and successfully solved for 1000 realizations with different initial conditions.

\subsection{Notation}\label{sec:Notation}
Let $||\boldsymbol{x}||:=\sqrt{\boldsymbol{x}^\top\boldsymbol{x}}$ be the Euclidean norm of a vector $\boldsymbol{x}$, and define the closed ball of dimension $d$ and radius $r$ as $B_r^d:=\{\boldsymbol{x} \in \reals^d:\|\boldsymbol{x}\| \leq r\}$, with $\mathcal{U}\!\left(B_d\right)$ the uniform distribution over $B_r^d$. Similarly, let $S_r^{d-1}:=\{\boldsymbol{x} \in \reals^d:\|\boldsymbol{x}\|=r\}$ denote the sphere of radius $r$, and $\mathcal{U}\!\left(S_r^{d-1}\right)$ the uniform distribution over $S_r^{d-1}$. For an arbitrary compact set $\Theta$, we define a projection operator $\Pi_{\Theta}(x):=\text{argmin}_{\theta\in\Theta}\|\theta-x\|$. For a vector $\boldsymbol{x}\in\reals^n$ we write $\boldsymbol{x}^{-i}:=\left[x^{1},\ldots,x^{i-1},x^{i+1},\ldots,x^{n}\right]^{T}$. Consider an inertial frame $\mathcal{C}:=\{c,\hat{\imath}_X,\hat{\imath}_Y,\hat{\imath}_Z\}$ fixed in the barycentre of the constellation, and three moving spacecraft frames $\mathcal{S}_i:=\{s_i,\hat{\imath}_x,\hat{\imath}_y,\hat{\imath}_z\}$, where $i\in I_s=\{1,2,3\}$, fixed in each spacecraft. We chose $\hat{\imath}_x$ pointing in the bisector of the two local lasers in their default configuration, $\hat{\imath}_z$ to be orthogonal to the plane spanned by the axes of the local lasers, and $\hat{\imath}_y$ completing the right-hand-rule.
\section{Problem Formulation}
An essential aspect of this problem is that the misalignments between laser beams are in the order of $\mu$rad--nrad. This fact can lead to numerical complications if the problem is not formulated properly. Employing the CCDs and QPDs as sensing devices is not only motivated by the high-precision requirements, but also by the fact that it enables the derivation of a \emph{relative} formulation that, in turn, will allow us to center our states around zero, reducing the scale of the units to $\mu$rad and avoiding numerical issues.

\subsection{System Model}\label{sec:system}
We start by considering three spacecraft with local moving frames $\mathcal{S}_i$, where $i\in I_s=\{1,2,3\}$, in triangular formation as represented in Fig. \ref{fig:geometry}. Each spacecraft has a state vector $\boldsymbol{x}^i:=\boldsymbol{x}^i(t)=[\alpha^i(t)\ \phi^i(t)\ \psi^i(t)\ \ell^i(t)]^\top\in\reals^4$ the first three elements denoting Euler angles that describe the inertial orientation of $\mathcal{S}_i$, and an element $\ell^i(t)$ describing the angle between both local lasers. For convenience, we consider the constellation frame $\mathcal{C}$ as inertial. It follows naturally to define the initial conditions of the spacecraft $\boldsymbol{x}_0^i:=\boldsymbol{x}^i(0)=[\alpha_0^i\ \phi_0^i\ \psi_0^i\ \ell_0^i]^\top\in\reals^4$, at the beginning of the laser acquisition phase. In order to adhere to a relative state formulation, the following relation becomes practical.
\begin{equation}\label{eq:dphi}
    \boldsymbol{x}^i=\boldsymbol{x}^i_0+\delta\boldsymbol{x}^i,\ i\in I_s.
\end{equation}
where $\delta\boldsymbol{x}^i:=\delta\boldsymbol{x}^i(t)=[\delta\alpha^i(t)\ \delta\phi^i(t)\ \delta\psi^i(t)\ \delta\ell^i(t)]^\top\in\reals^4$ are small deviations from the initial condition. Let us group the three relative state vectors into a full state vector $\delta\boldsymbol{x}(t)=[\delta\boldsymbol{x}^{1\top}\ \delta\boldsymbol{x}^{2\top}\ \delta\boldsymbol{x}^{3\top}]^\top$, or just $\delta\boldsymbol{x}$, for compactness.
Determining a full model for $\delta\dot{\boldsymbol{x}}$ lies out of the scope of this paper. We will assume, nonetheless, that each spacecraft has its own internal controller with actuation resolution in the order of nanoradians. The spacecraft' states will then evolve according to their closed-loop dynamics denoted by the function $f$,
\begin{equation}\label{eq:phidot}
    \delta\dot{\boldsymbol x}=f(\delta\boldsymbol{x},\boldsymbol{u}_{k-1}),\ t\in[t_{k-1},t_{k}).
\end{equation}
which has piecewise-constant reference vectors $\boldsymbol{u}_k:=\boldsymbol{u}(t_k)=[\boldsymbol{u}_k^{1\top}\ \boldsymbol{u}_k^{2\top}\ \boldsymbol{u}_k^{3\top}]^\top$, with $\boldsymbol{u}_k^i:=\boldsymbol{u}^i(t_k)\in U^i\subseteq\reals^4$, as inputs. Using a constraint box for the references with $b\in\reals$, we have that
\begin{equation}\label{eq:Uset}
U^{i}=[-b,b]^4,\quad \forall i\in I_s,
\end{equation}
Considering the system behaviour under the closed-loop dynamics presented above, we make the following fair assumption about the internal controller.
\begin{assumption}\label{sec:ass1}
    \emph{Taking the system \eqref{eq:phidot}, we assume that the states $\delta\boldsymbol{x}$ at time $t\in[t_{k-1},t_{k})$ will converge to the reference $\boldsymbol{u}_{k-1}$ with a constant time $\tau_c\in\reals_{\geq 0}$ such that,
\begin{equation}\label{eq:cl_convergence}
    \lim_{t\rightarrow t_{k}} ||\delta\boldsymbol{x}(t)-\boldsymbol{u}_{k-1}||\leq \exp\left(-\frac{(t-t_k)}{\tau_c}\right).
\end{equation} }
\end{assumption}
Additionally, we build the guidance system so that it provides references at a slower pace than the closed-loop convergence rate \eqref{eq:cl_convergence}, which supports the next assumption.
\begin{assumption}\label{sec:ass2}
    \emph{Let the guidance system provide reference samples every time period of $\tau_g:=t_{k}-t_{k-1}\in\reals_{\geq 0}$ with $\tau_g\gg\tau_c$. Then, if Assumption \ref{sec:ass1} holds, it follows that
\begin{equation}
    ||\delta\boldsymbol{x}(t_k)-\boldsymbol{u}_{k-1}||\leq \exp\left(-\frac{\tau_g}{\tau_c}\right),
\end{equation}
    which indicates that $\delta\boldsymbol{x}_k\approx\boldsymbol{u}_{k-1}$, where $\delta\boldsymbol{x}_k:=\delta\boldsymbol{x}(t_k)$.}
\end{assumption}

\subsection{Objective Functions}
Let us introduce the objective functions $h^i\!:\!\boldsymbol{U}\!\rightarrow\!\reals_{\geq 0}$ of $\boldsymbol{u}$ that each spacecraft $\mathcal{S}_i$ aims to minimize, where $\boldsymbol{U}=U^1\times U^2\times U^3$ denotes the set of joint references. Each function $h^i$ penalizes the total misalignment measured by both detectors on each spacecraft. Provided that all spacecraft receive two beam pulses, then $h^i$ is designed to be zero only when the misalignment between both local sensors and their received  beams is also zero. We define another function $y_{ij}:U^i\times U^j\!\rightarrow\!\reals_{\geq 0}$ as a model for the individual misalignment measured by each individual sensor. Given the relation between $\delta\boldsymbol{x}_k$ and $\boldsymbol{u}_{k-1}$ from Assumption \ref{sec:ass2}, we propose the objective function to be
\begin{equation}\label{eq:h}
    h^i(\boldsymbol{u}_{k-1})=\!\!\sum_{j\in I_s^{ij}}\!w y^2_{ij}(\boldsymbol{u}_{k-1}^i,\boldsymbol{u}_{k-1}^j),\ \forall i\in I_s.
\end{equation}
We normalize the objectives with a scalar weight $w\in\reals$ using the worst achievable value of $h^i$ to avoid numerical issues. Herein, the worst misalignment corresponds to the size of the UC, i.e. 9$\mu$rad.

Note that this setup is clearly a cooperative game that will be denoted as $\Gamma(3,\{h^i\},\{U^i\})$, given that each spacecraft can be deemed an agent trying to minimize a common objective, i.e. the sum of all the individual objectives $h^{i}$. However, any arbitrary spacecraft $\mathcal{S}_{i}$ can only observe its local objective $h^{i}$ by means of its own sensors. Moreover, any reference $\boldsymbol{u}^i\in U^i$ selected by this spacecraft will have an impact on the objectives of the other two spacecraft.

We proceed to determine an expression for $y_{ij}$ that rely first on the relative states $\delta\boldsymbol{x}$, and later on $\boldsymbol{u}$. To that end, let $\bar{\boldsymbol{x}}^i\in\reals^4$ be some desired inertial state to be attained by each spacecraft and its maneuverable telescope, for which all the lasers are perfectly aligned.
\begin{remark}\label{sec:remark1}
    \emph{Note that the target attitude vectors $\bar{\boldsymbol{x}}$ depend on the actual position and orientation of all the spacecrafts. Errors in these quantities are unknown to each spacecraft and yield to non-equilateral triangle formations.}
\end{remark}
It follows to define the initial error, $\delta\boldsymbol{x}^i_0\in\reals^4$, of spacecraft $\mathcal{S}_i$ from its target $\bar{\boldsymbol{x}}^i$.
\begin{equation}\label{eq:dphi_0}
    \delta\boldsymbol{x}^i_0=\bar{\boldsymbol{x}}^i-\boldsymbol{x}^i_0.
\end{equation}
It is evident, then, that both quantities, $\bar{\boldsymbol{x}}^i$ and $\delta\boldsymbol{x}^i_0$, are unknowns. Additionally, notice that in order to achieve such a triangular formation, the difference between the relative states of any pair of spacecraft $\mathcal{S}_i$ and $\mathcal{S}_j$, with $(i,j)\in \{i,j\in I_s:i\neq j\}$, has to result in a constant, but unknown, angle vector $\bar{\boldsymbol{x}}_{ij}\in\reals^4$.
\begin{equation}\label{eq:phi^ij}
    \bar{\boldsymbol{x}}_{ij}=\bar{\boldsymbol{x}}^i-\bar{\boldsymbol{x}}^j,
\end{equation}
which value depends on the spacecraft pair involved.
\begin{remark}
    \emph{Since the triangular formation might not be equilateral, see Remark \ref{sec:remark1}, the relative vector $\bar{\boldsymbol{x}}_{ij}$ is unknown.}
\end{remark}
Based on the previous equations, we propose the following function that models the misalignment of the beam emitted by $\mathcal{S}_j$ and measured by $\mathcal{S}_i$ at discrete time $t_k$.
\begin{equation}\label{eq:y}
    y_{ij}(\boldsymbol{x}_k^i, \boldsymbol{x}_k^j) = ||(\boldsymbol{x}_k^i-\boldsymbol{x}_k^j)-\bar{\boldsymbol{x}}_{ij}||,
\end{equation}
As expected, the measured misalignment $y_{ij}$ is zero only when the difference between the inertial states of the spacecraft pair is equal to the target vector $\bar{\boldsymbol{x}}_{ij}$, and positive otherwise. Saturation on $y_{ij}$ when it exceeds 9$\mu$rad is disregarded, since it would be required to switch to another guidance scheme governed by other sensors if the states were to overshoot outside of the UC. We continue to modify \eqref{eq:y} due to its dependency on inertial variables. Replacing \eqref{eq:dphi}, \eqref{eq:dphi_0} and \eqref{eq:phi^ij} to express $y_{ij}$ in terms of relative states, leads to the final measurement equation,
    $$y_{ij} (\delta\boldsymbol{x}_k^i, \delta\boldsymbol{x}_k^j)= ||(\delta\boldsymbol{x}_k^i - \delta\boldsymbol{x}_k^j)-(\delta\boldsymbol{x}^i_0 - \delta\boldsymbol{x}^j_0)||,$$
which according to Assumption \ref{sec:ass2},
\begin{equation}\label{eq:yij}
    \begin{aligned}
        y_{ij} (\delta\boldsymbol{x}_k^i, \delta\boldsymbol{x}_k^j)&\approx y_{ij} (\boldsymbol{u}_{k-1}^i, \boldsymbol{u}_{k-1}^j)\\
                    & =||(\boldsymbol{u}_{k-1}^i - \boldsymbol{u}_{k-1}^j)-(\delta\boldsymbol{x}^i_0 - \delta\boldsymbol{x}^j_0)||,
    \end{aligned}
\end{equation}

Observe that, in practice, we only obtain values for the measurement $y_{ij}$, and the model \eqref{eq:yij} is just intended for simulation purposes.

This concludes the description of the game $\Gamma(3,\{h^i\},\{U^i\})$.


\section{Extremum Seeking}
Throughout this section, we will use the words spacecraft and players indistinctly. First, we will show that the game $\Gamma(3,\{h^i\},\{U^i\})$ has at least one \emph{Nash Equilibrium} (NE), i.e. a stable solution $\boldsymbol{u}^\ast$ which satisfies
\begin{equation}
    h^i(\boldsymbol{u}^{i\ast},\boldsymbol{u}^{-i\ast})\leq h^i(\boldsymbol{u}^{i},\boldsymbol{u}^{-i\ast}),\quad\boldsymbol{u}^i\in U^i,\ \forall i\in I_s.
\end{equation}
Afterwards, we will present a payoff-based strategy in which each spacecraft $\mathcal{S}_i$ at time $t_{k}$ computes an update $\boldsymbol{u}^{i}_{k}$ based on their most recent reference, $\boldsymbol{u}^{i}_{k-1}$, and the value of their objective, $h^{i}(\boldsymbol{u}_{k-1})$.

\subsection{Existence of Nash Equilibria}
To show the existence of a NE in the game $\Gamma(3,\{h^i\},\{U^i\})$, we first put forward the following theorem.
\begin{theorem}
\label{thm:Theonlytheoreminpaper}
    (Proposition 2.2.9 in \cite{facchinei2003finite}) \emph{Consider a game $\Gamma(N,\{q^i\},\{A^i\})$ with $N\in\mathbb{Z}_{>0}$. Let $\boldsymbol{a}^i\in A^{i}$, with $A^i\subset\reals^{n_i}$ be compact and convex, and $q^{i}$ be continuously differentiable. If each function $q^{i}(\boldsymbol{a}^{i},\boldsymbol{a}^{-i})$ is convex in $\boldsymbol{a}^{i}$ for every fixed $\boldsymbol{a}^{-i}$, then the set of Nash equilibria is nonempty and compact.}
\end{theorem}
In the game $\Gamma(3,\{h^i\},\{U^i\})$ studied in this paper, we have $U^{i}\subset \reals^{4}$, defined as \eqref{eq:Uset}, being convex and compact. Furthermore, all functions $h^{i}$ are continuously differentiable, and $\forall i\in I_{s}$ it holds that $h^{i}\left(\boldsymbol{u}^{i},\boldsymbol{u}^{-i}\right)$ is convex in $\boldsymbol{u}^i$ for a fixed $\boldsymbol{u}^{-i}$, with $y_{ij}$ modeled as in \eqref{eq:yij}. Therefore, Theorem \ref{thm:Theonlytheoreminpaper} states that the set of Nash equilibria for the game $\Gamma(3,\{h^i\},\{U^i\})$ is nonempty and compact.

\subsection{Learning of Nash Equilibria}
In order to seek a NE, we propose a strategy inspired by the work of \cite{Tatjana_alg_2019}. In their work, a gradient estimate of the objectives of each player is obtained using a randomization technique. We assume that at time $t_{k}$ each spacecraft randomly selects --or \emph{mixes}-- their strategy according to the following
\begin{equation}
\label{eq:randomization}
    \boldsymbol{u}^{i}_{k} = \boldsymbol{\mu}^{i}_{k} + r_{k}\boldsymbol{\zeta}^{i}_{k},\quad \boldsymbol{\zeta}^{i}_{k}\sim \mathcal{U}\left(S^{3}_{1}\right),
\end{equation}
where $\boldsymbol{\mu}^{i}_{k}\in\reals^4$ is the mean values of the reference $\boldsymbol{u}^{i}_{k}$, with an update equation that will be discussed later in this section, and $r_{k}\in\reals$ is a time varying exploration parameter.
Afterwards, each spacecraft obtains a gradient estimate $\boldsymbol{g}^{i}_{k}\in\reals^4$ based on the residual-feedback method introduced first in \cite{Res_feedback_2022} as follows
\begin{equation}
\label{eq:grad_est}
    \boldsymbol{g}^{i}_{k} = \frac{4}{r_{k}}\left(h^{i}(\boldsymbol{u}_{k})-h^{i}(\boldsymbol{u}_{k-1})\right)\boldsymbol{\zeta}^{i}_{k}.
\end{equation}
One of the key properties of $\boldsymbol{g}^{i}_{k}$ is the fact that
$$\mathbb{E}_{\boldsymbol{\zeta}^{i}_{k}}\!\!\left[\boldsymbol{g}^{i}_{k}\right]=\frac{\partial\tilde{h}^{i}(\boldsymbol{\mu}_{k})}{\!\partial\boldsymbol{\mu}_k^i},$$
where $\tilde{h}^{i}$ is a smooth version of $h^{i}$ obtained by averaging the function $h^{i}$ with a uniform random variable $\boldsymbol{\xi}_k$ over the unit ball. That is
$$\tilde{h}^{i}(\boldsymbol{\mu}_{k}):=\mathbb{E}_{\boldsymbol{\xi}_{k}}\![h^i(\boldsymbol{\mu}_{k}+r_{k}\boldsymbol{\xi}_{k})],\quad\boldsymbol{\xi}_{k} \sim \operatorname{\mathcal{U}}\left(B^{12}_1\right),$$ which is a standard result in zeroth-order optimization (see \cite{chen2022improve,Flaxman_2005_zo}). In other words, the value $\boldsymbol{g}^{i}_{k}$ is an unbiased estimator for the gradient of the smoothed version $\tilde{h}^{i}$. Using two function evaluations for the estimator $\boldsymbol{g}^{i}_{k}$ in \eqref{eq:grad_est} was shown by \cite{Res_feedback_2022} to reduce the variance of the estimates, which provides faster convergence than the case of using one function evaluation as done by \cite{Tatjana_alg_2019}.

After introducing the gradient estimate $\boldsymbol{g}^{i}_{k}$, the reference $\boldsymbol{u}^{i}_{k}$ for each spacecraft is calculated using \eqref{eq:randomization} in which $\boldsymbol{\mu}^{i}_{k}$ is determined according to
\begin{equation}
    \label{eq:Update_eq}
    \boldsymbol{\mu}^{i}_{k} = \Pi_{\tilde{U}_k^{i}}\left(\boldsymbol{\mu}^{i}_{k-1} - \eta_{k-1}^{i}\boldsymbol{g}^{i}_{k-1} + \rho\left(\boldsymbol{\mu}^{i}_{k-1}-\boldsymbol{\mu}^{i}_{k-2}\right)\right),
\end{equation}
where $\eta_{k}\in\reals_{>0}$ is a time-varying step size and the term $\rho\left(\boldsymbol{\mu}^{i}_{k-1}-\boldsymbol{\mu}^{i}_{k-2}\right)$ represents a momentum term with parameter $\rho\in[0,1)$. Our introduction of the momentum term in \eqref{eq:Update_eq} leads to a faster convergence for the seeking scheme to a NE. See \cite{qian1999momentum} for additional details regarding momentum in gradient descent algorithms. The set $\tilde{U}^{i}_{k}=\left[-b^{i}_{k}~b^{i}_{k}\right]^{4}$ in \eqref{eq:Update_eq} is time-dependent, with bounds $$b^{i}_{k}=\beta^{k}b^{i}_{0}+(1-\beta^k)b,\quad 0\leq\beta<1,$$ where $b_0\in\reals$ is the initial size chosen for the box constraint and $b$ the final, coinciding with the original constraint in \eqref{eq:Uset}. This time dependency on the box constraints is introduced to help avoiding violent changes in the references at the beginning of the phase, which can cause overshooting outside the UC. Note that we also propose a uniform distribution for $\boldsymbol{\zeta}^{i}_{k}$, instead of a normal distribution as done by \cite{Tatjana_alg_2019}, since it has a finite support as opposed to the normal distribution. This ensures the boundedness of $\boldsymbol{u}$ which is more suitable for the restriction \eqref{eq:Uset} to hold. The summary of the seeking strategy used by each spacecraft $\mathcal{S}_i$ is provided in Algorithm \ref{alg:NEseekings}.
\begin{remark}
    \emph{It is easy to check that $\Gamma(3,\{h^i\},\{U^i\})$ is a convex and monotone game. The structure present by n-player convex and monotone games rules out the existence of non-Nash equilibria (\cite{rosen1965existence}) and thus, a gradient-play strategy will converge to a NE (\cite{mazumdar2020gradient}). Moreover, gradient-bandit strategies have been proposed and shown to converge to a NE in monotone games, see \cite{Tatjana_alg_2019} and \cite{Gao2}. In this paper, we demonstrate empirical convergence of our gradient-bandit algorithm to a NE. Yet, a theoretical analysis and proof are necessary for a complete understanding of the algorithm's behavior.}
\end{remark}
\vspace{-1mm}
\begin{algorithm}[h]
\DontPrintSemicolon
\Parameter{$\gamma_{r}>0,~a_{r}>0$ such that $r_{k}=\gamma_{r}/(k)^{a_{r}}$, $\gamma_{\eta}>0,~a_{\eta}>0$ such that $\eta_{k} = \gamma_{\eta}/(k)^{a_{\eta}}$, $0\leq\rho<1,~0\leq\beta<1,$ and $\varepsilon>0$.}
  \KwInput{initial $\boldsymbol{\mu}^{i}_{0}=\boldsymbol{\mu}^{i}_{-1}=\boldsymbol{0}$.}
  \KwOutput{$\boldsymbol{u}^{i}_{k}$}
  $k=0$\;
  \While{$\left\|h^i(\boldsymbol{u}_{k})-h^i(\boldsymbol{u}_{k-1})\right\|\geq \varepsilon$}{
  $k \leftarrow k+1$,\;
    Obtain the measurements $y_{ij},~\forall j\in I_{s}^{ij}$ and compute $h^{i}\left(\boldsymbol{u}_{k-1}\right)$ from \eqref{eq:h}.\;
  Compute $\boldsymbol{g}^{i}_{k-1}$ using \eqref{eq:grad_est}.\;
  Compute $\boldsymbol{\mu}^{i}_{k}$ using \eqref{eq:Update_eq}.\;
  Compute $\boldsymbol{u}^{i}_{k}$ using \eqref{eq:randomization}.\;}
  Set $\boldsymbol{u}^{i}_{k}=\boldsymbol{\mu}^{i}_{k}$.
    \caption{NE seeking for spacecraft $\mathcal{S}_i, i\in I_{s}$}
\label{alg:NEseekings}
\end{algorithm}

\begin{figure*}[ht]
    \centering
    \includegraphics[width=1\textwidth]{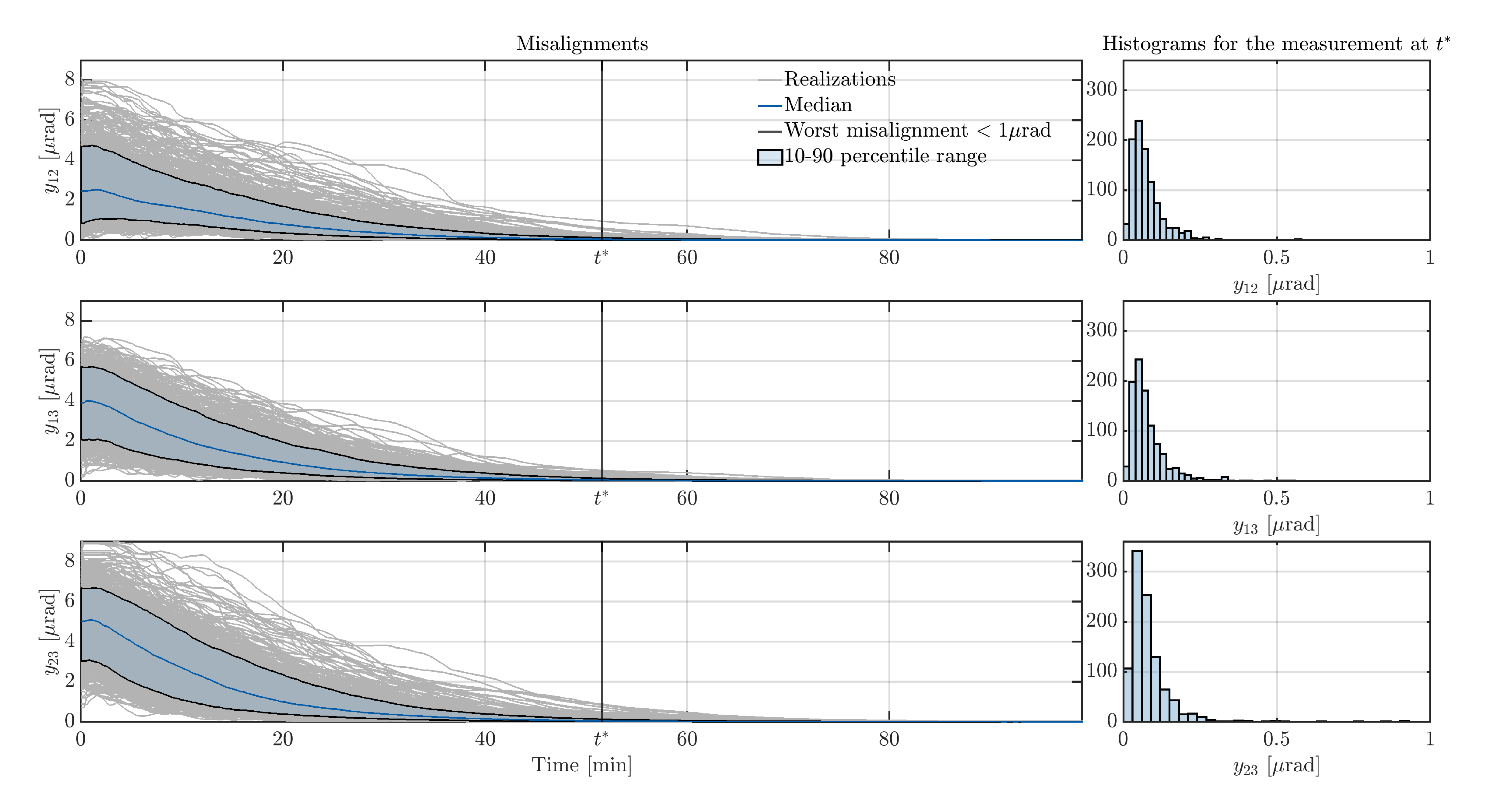}
    \caption{Simulation of 1000 realizations with different initial conditions inside the UC. Each realization is simulated for 5000 iterations, and $t^\ast$ marks the time at which the worst realization reaches the field of view of the QPD, i.e. all misalignments are bellow $1\mu$rad. The histograms illustrates the misalignments measured at time $t^\ast$.}
    \label{fig:MCresults}
\end{figure*}
\section{Simulation results}
The following is a description of the simulation setup\footnote{Code of Algorithm \ref{alg:NEseekings} available at \url{https://github.com/aitor-rg/LISA-pointing-game}} to test the performance of Algorithm \ref{alg:NEseekings}.

All three spacecraft lie inside an UC of $9\mu$rad, and thus we randomize the initial conditions $\delta\boldsymbol{x}_0$ such that between any spacecraft pair it holds that $||\delta\boldsymbol{x}_0^i-\delta\boldsymbol{x}_0^j||\leq9\mu$rad. In accordance with Assumption \ref{sec:ass1}, the states have been propagated using a first-order system
\begin{equation}\label{eq:ode}
    \delta\dot{\boldsymbol{x}}(t)=\frac{1}{\tau_c}(\boldsymbol{u}_{k-1}-\delta\boldsymbol{x}(t)),\quad t\in[t_{k-1},t_k).
\end{equation}
The convergence time of the controller is defined to be $\tau_c=1$s. Given that optical measurements are spaced in time by 16 seconds --due to the interferometer arm length-- then the guidance system can provide reference updates every $\tau_g=\tau_c+16=17$s. For this time rate, it is fair to consider that $\tau_g\gg\tau_c$ and, thus, that Assumption \ref{sec:ass2} also holds.

A total of 1000 realizations have been drawn for the initial conditions $\delta\boldsymbol{x}_0$. Instead of picking a value for the stopping parameter $\varepsilon$ in line 2 of the algorithm, each of the initial states have been numerically propagated for $K=5000$ iterations using \eqref{eq:ode}. At each iteration $k$, Algorithm \ref{alg:NEseekings} is used to update the references $\boldsymbol{u}_k$, corresponding to time $t_k=k\tau_g$. For convenience, a time increment $\Delta t=\tau_g$ has been chosen, so that a new reference update is computed at each iteration. Equivalently, we have $\Delta t=\tau_g=1$ and $\tau_c=1/\Delta t=1/17$. The initial bound for the time-dependent constraint $\tilde{U}_k^i$ is calculated from the initial conditions, $\boldsymbol{u}_0=\mathbf{0}$, as $$b^i_0 = 0.3(1-h^i(\mathbf{0})).$$ The remaining parameters required for the algorithm are selected as follow: $a_{r}=0.2,~a_{\eta} = 0.5,~\gamma_{r}=0.58,~\gamma_{\eta}=4.5,~\rho = 0.93,~\beta = 0.01,~w=1/9^2$ and $b=4.5$.
\begin{table}[h]
    \centering
    \caption{Average misalignments at time $t^\ast$, the time when the worst realization is below $1\mu$rad, using Algorithm \ref{alg:NEseekings} and Algorithm \ref{alg:NEseekings}$^\dagger$.}
    \begin{tabular}{|c|c|c|}
        \hline
        Misalignment &  Algorithm \ref{alg:NEseekings} & Algorithm \ref{alg:NEseekings}$^\dagger$ \\
        \hline
         $y_{12}~[\mu\text{rad}]$ & 0.0794 &  4.10\\
         \hline
         $y_{13}~[\mu\text{rad}]$ &  0.0804 &  4.25\\
         \hline
         $y_{23}~[\mu\text{rad}]$  & 0.0843 & 4.54\\
         \hline
    \end{tabular}
    \label{tab:Meany}
\end{table}

The results of the Monte Carlo simulation are displayed in two fashion. In Fig. \ref{fig:MCresults}, we show the lasers misalignments between spacecraft, $y_{ij}$, for each realization. The robustness of the algorithm against different initial conditions is exhibited, by showing the convergence of each trajectory as a function of time, and their settling misalignment at time $t^\ast$ in a histogram. Time $t^\ast$ correspond to the time at which the worst realization enters the field of view of the QPD, and thus, all misalignments are below $1\mu$rad. We also show the 10-90 percentile range of the realizations in the blue shaded area, and the median in darker blue. Additionally, Table \ref{tab:Meany} provides details of the mean value of the misalignments at $t^\ast$. Moreover, these values are compared to the results obtained using a version of Algorithm \ref{alg:NEseekings} without momentum and residual-feedback terms. We refer to this as Algorithm \ref{alg:NEseekings}$^\dagger$. Note that for such a case, the results are severally degraded in accuracy if simulated for the same period of time.

Finally, the time taken to converge inside the field of view of the QPD, $<1\mu$rad, for the worst case is 51.57min.

\section{Conclusion}
This work presents a guidance scheme for the LISA pointing acquisition phase, which commands three spacecraft at once to establish interspacecraft laser links for GWs detection. The guidance scheme is rooted in notions of game theory, which have been extended to deliver the main contribution of the paper, Algorithm \ref{alg:NEseekings}, i.e. a new algorithm for learning of Nash Equilibria in cooperative games. The convergence of this algorithm to a NE is shown empirically and future work is needed to establish a rigorous proof of the algorithm's convergence properties.

The proposed methodology is based on the idea of defocusing the laser beam which, as opposed to the main scanning strategies in the literature, do not rely on ground communication during the acquisition process. It is a fully autonomous system, and it does not require changes or additional mechanisms. Each spacecraft is assumed to have micro-thrusters providing three rotational degrees of freedom, and an extra degree of freedom is considered for one of the two telescopes. Simulation results show that the algorithm is robust against initial conditions lying inside the uncertainty cone, and the worst misalignment between the laser and the interferometer detectors is $<1\mu$rad. This means that the laser still is inside the field of view of the quadrant photodiode, which provide good margins of error. The time duration of the pointing acquisition --reaching below $1\mu$rad-- for the worst simulated case is 51.57min, below 1 hour.

Moreover, a new idea has been put forward on how to manipulate the lasers during this phase.  It is suggested that the lasers should emit pulses, rather than a continuous wave, in order to enable the emission and reception of optical information simultaneously between the three spacecraft. However, more technical work is needed on the laser front to ensure sufficient signal to noise ratio when defocusing the laser beam, as well as when emitting optical pulses. A certain scope for future work is the inclusion of a full dynamical model of the spacecraft with disturbances.

\bibliography{ifacconf}             








\end{document}